\documentclass[12pt,a4paper ]{article}
\usepackage{graphicx}

\input epsf

\newcommand{\frat}[2]{\frac{\textstyle #1}{\textstyle #2}}
\newcommand{\vf}[1]{\mbox{\boldmath $#1$}}

\begin{document}

\begin{center}
{\Large \bf  Limits of colour charge interaction in the instanton liquid}\\
\vspace{0.5cm} S.V. Molodtsov$^{1,2}$, G.M. Zinovjev$^3$\\
\vspace{0.5cm} $^1${\small\it State Research Center,
Institute of Theoretical and Experimental Physics,
117259, Moscow, RUSSIA}\\
$^2${\small\it Joint Institute for Nuclear Research,
RU-141980 BLTP JINR, Dubna, RUSSIA}\\
$^3$ {\small \it Bogolyubov Institute for Theoretical Physics,\\
National Academy of Sciences of Ukraine, UA-03143, Kiev, UKRAINE}
\end{center}
\vspace{0.5cm}

\begin{center}
\begin{tabular}{p{16cm}}
{\small{The effect of external colour field impact on the instanton liquid
is studied. In the course of this study the corresponding effective Lagrangians
are derived for both regimes of weak and strong external field and in long wave-length
approximation. The example of Euclidean colour point-like source is analyzed in
detail and the feedback of field on the instanton liquid is estimated as a
function of source intensity.}}
\end{tabular}
\end{center}
\vspace{0.5cm}

The declarations of discovering new state(s) of matter in relativistic
heavy ion collisions at RHIC which are actively wandering in the papers nowadays
are sometimes based on the results of different nature. From one side it is
the striking result of direct experimental measurements of a strong suppression
(comparing to $pp$ and $pA$ collisions)
of particle production at high transverse momentum well-known as a jet
quenching. And although the jet reconstruction in these experiments is a
nontrivial task the (and accompanying) result(s) is(are) interpreted as a
degradation of hard parton (initiating a jet) energy induced by medium
(new thermalized matter) produced in collision long before hadronizing in the
QCD vacuum. On the other hand the convincing success of phenomenological
analysis of the other measurable characteristics  based on the perfect liquid hydrodynamics
results in the questions about the sort of plasma (if produced) and the
origin of the QCD vacuum Ref.\cite{Barb}.

In the present paper we are trying to understand if the instanton liquid
model of QCD vacuum could be indicative in interpreting these RHIC results.
The instanton liquid here is considered in the framework of the stochastic ensemble of
instantons in the singular gauge and the generating functional is evaluated
by the variational principle developed in the paper Ref.\cite{DP}. Comparative simplicity of the
superposition ansatz and variational procedure allows us to analyze  the effects almost
analytically.

As a major configuration saturating the generating functional
$$Z=\int D[A]~e^{-S(A)}~$$ where $S(A)$ is a standard Yang-Mills action we take
the approximate solution for the Yang-Mills equations in the form of the
following ~superposition
\begin{equation}
\label{1} {\cal  A}^{a}_\mu(x)=B^a_{\mu}(x)+\sum_{i=1}^N
A^{a}_\mu(x;\gamma_i)~,
\end{equation}
here $A^a_{\mu}$ implies the field of (anti-)instantons in the singular gauge
\begin{equation}
\label{2}
A^a_{\mu}(x)=\frat2g~\omega^{ab}\bar\eta_{b\mu\nu}~a_\nu(y)~,~~~
a_\nu(y)= \frat{\rho^2}{y^2+\rho^2}~\frat{y_\nu}{y^2}~,~~~y=x-z~,
\end{equation}
with the parameters $\gamma_i=(\rho_i,z_i,\omega_i)$ describing the $i$-th
instanton of the $\rho$ size centered at the pseudo-particle coordinate $z$, with the matrix
of colour orientation $\omega$, and $g$ denotes the coupling constant of
non-abelian field. For the anti-instanton the 't Hooft symbols should be changed according
to $\bar\eta \to \eta$, and the external field $B^{a}_{\mu}(x)$ is viewed as well
as the pseudo-particle field in quasiclassical approximation.

The external field interaction with an individual pseudo-particle is defined by
the following strength tensor
\begin{equation}
\label{3}
G_{\mu\nu}^a=G_{\mu\nu}^a(B)+G_{\mu\nu}^a(A)+G_{\mu\nu}^a(A,B)~,
\end{equation}
where the first two terms correspond to standard strength tensors of
non-abelian field
\begin{equation}
\label{4}
 G_{\mu\nu}^a(A)=\partial_\mu  A^a_{\nu}-\partial_\nu
 A^a_{\mu}+g~f^{abc} A^b_{\mu} A^c_{\nu}~,
\end{equation}
with entirely anti-symmetric tensor $f^{abc}$. In particular,
\begin{equation}
\label{5}
G^a_{\mu\nu}(A)=-\frat4g~\omega^{ak}\bar\eta_{k\alpha\beta}~
M_{\mu\alpha}M_{\nu\beta}~\frat{\rho^2}{(y^2+\rho^2)^2}~,
\end{equation}
where $M_{\mu\nu}=\delta_{\mu\nu}-2~\hat y_\mu \hat y_\nu,~ \hat
y_\mu=\frac{y_\mu}{|y|}$.
The 'mixed' component of the instanton strength field looks like
\begin{equation}
\label{6}
G_{\mu\nu}^a(A,B)=g~f^{abc}(B^b_{\mu}A^{c}_\nu-B^b_{\nu}A^{c}_\mu)
=g~f^{abc}\omega^{cd}~\frat2g~ (B^{b}_\mu
~\bar\eta_{d\nu\alpha}-B^{b}_\nu
~\bar\eta_{d\mu\alpha})~a_{\alpha}(y)~.
\end{equation}
Calculating now $G^2$ we receive the partial contributions of external field
and each separate pseudo-particle as
\begin{eqnarray}
\label{7}
G_{\mu\nu}^a~ G_{\mu\nu}^a &=&
G_{\mu\nu}^a(B)~G_{\mu\nu}^a(B) + G_{\mu\nu}^a(A)~G_{\mu\nu}^a(A) +
G_{\mu\nu}^a(A,B)~G_{\mu\nu}^a(A,B) +\nonumber\\ [-.2cm]
\\[-.25cm]
&+&2~ G_{\mu\nu}^a(B)~G_{\mu\nu}^a(A) + 2~
G_{\mu\nu}^a(B)~G_{\mu\nu}^a(A,B) + 2~
G_{\mu\nu}^a(A)~G_{\mu\nu}^a(A,B)~, \nonumber
\end{eqnarray}

In order to keep the further steps as simple and transparent as possible we
limit ourselves with the standard sum of partial contributions in the
superposition ansatz action and hold the highest in IL density (precisely in
packing fraction parameter $n\rho^4$) one particle contributions
\begin{equation}
\label{8}
S(B,\gamma)=\int dx~\frat{G_{\mu\nu}^a~ G_{\mu\nu}^a}{4}\simeq \sum_i \int dx ~
\frat{G_{\mu\nu}^a(i)~ G_{\mu\nu}^a(i)}{4}~.
\end{equation}
The crossing terms of different pseudo-particles (which are
proportional to the IL density squared) are neglected here
because of very small packing fraction parameter characteristic to
IL, i.e.  $n\rho^4\sim 0.01$. Thus, the regularized generating
functional for the IL model takes the following form (for
denotations see Ref.\cite{DP})
\begin{equation}
\label{9} Y=\int D[B]~ \frat{1}{N!} \int \prod_{i=1}^N~ d
\gamma_i~~e^{-S(B,\gamma)}~.
\end{equation}
 The analysis can be easily done for the weak external field assuming the
characteristic IL parameters, averaged pseudo-particle size $\bar \rho$ and
the IL density $n$, unchanged and coinciding with their vacuum magnitudes. Those
are fixed by some repulsive mechanism (see, however, the remark at the end of
paper) for the particular choice of saturating configuration done above. In this case,
from now on the integration over pseudo-particle sizes becomes unessential and all the
pseudo-particles might be considered as having the same size $\bar\rho$. It
allows us to average over the pseudo-partile positions and colour orientations
only by calculating the generating functional (\ref{9}). Making use the cluster
decomposition we obtain the corresponding average of exponential as
\begin{equation}
\label{10}
 \langle \exp (-S)\rangle_{\omega z}=\exp\left(~\sum_k
\frat{(-1)^k}{k!}~\langle\langle S^k\rangle\rangle_{\omega z}\right)~,
\end{equation}
where $\langle S_1\rangle=\langle\langle S_1\rangle\rangle$,
 $\langle S_1 S_2\rangle=\langle S_1\rangle\langle S_2\rangle
+\langle\langle S_1 S_2\rangle\rangle, \dots$.\\
The first cumulant is simply defined by the action averaged. Taking into
account the direct form of field strength tensors (\ref{5}) and (\ref{6}) it is evident
that the following terms will only be present in the partial contribution
after averaging over colour orientation
$$ \langle G_{\mu\nu}^a~ G_{\mu\nu}^a \rangle_\omega=
G_{\mu\nu}^a(B)~G_{\mu\nu}^a(B) +\langle
G_{\mu\nu}^a(A)~G_{\mu\nu}^a(A)\rangle_\omega +
\langle G_{\mu\nu}^a(A,B)~G_{\mu\nu}^a(A,B) \rangle_\omega
 + 2~\langle G_{\mu\nu}^a(A)~G_{\mu\nu}^a(A,B)\rangle_\omega~.$$
Performing the colour averaging we use the equality
\begin{equation}
\label{12}
\langle
\omega^{ak}\omega^{cd}\rangle=\frat{\delta^{ac}\delta^{kd}}{N_c^{2}-1}~,
\end{equation}
implying $N_c$ as the number of colours. As a result we have that the last term
$$ 2~G_{\mu\nu}^a(A)~G_{\mu\nu}^a(A,B)=-\frat{16}{g} \omega^{ak}~
\bar\eta_{k\alpha\beta}~M_{\mu\alpha}M_{\nu\beta}~\frat{\rho^2}{(y^2+\rho^2)^2}~
f^{abc}\omega^{cd}~ (B^{b}_\mu ~\bar\eta_{d\nu\alpha}-B^{b}_\nu~
\bar\eta_{d\mu\alpha})~a_{\alpha}(y)~$$
disappears due to the antisymmetric tensor $f^{abc}$,
$$\langle G_{\mu\nu}^a(A)~G_{\mu\nu}^a(A,B)\rangle_\omega=0~. $$
Analyzing now the contribution of 'mixed' (repulsive) component
$$G_{\mu\nu}^a(A,B)~G_{\mu\nu}^a(A,B)=4~f^{abc}\omega^{cd}~
(B^{b}_\mu~ \bar\eta_{d\nu\alpha}-B^{b}_\nu~
\bar\eta_{d\mu\alpha})~a_{\alpha}(y)
~f^{akm}\omega^{mn}~(B^{k}_\mu ~\bar\eta_{n\nu\gamma}-
B^{k}_\nu ~\bar\eta_{n\mu\gamma})~a_{\gamma}(y)~,$$
and averaging Eq.(\ref{12}) over the colour orientation (making use another
equality)
$$f^{abc}f^{akc}=N_c~\delta^{bk}~,$$
we have
\begin{equation}
\label{13}
\langle G_{\mu\nu}^a(A,B)~G_{\mu\nu}^a(A,B)
\rangle_\omega=\frat{4~N_c}{N_c^{2}-1}~
(B^{b}_\mu~
\bar\eta_{d\nu\alpha}-B^{b}_\nu~\bar\eta_{n\mu\alpha})~a_{\alpha}(y)~
(B^{b}_\mu
~\bar\eta_{d\nu\gamma}-B^{b}_\nu~\bar\eta_{d\mu\gamma})~a_{\gamma}(y)~.
\end{equation}
Averaging over the pseudo-particle positions results in the following integral
$$\int
\frat{dz}{V}~a_{\alpha}(y)~a_{\gamma}(y)=\delta_{\alpha\gamma}~\frat{I}{V}~,$$
where
\begin{equation}
\label{14}
I=\frat{\pi^2}{4}~\rho^2~,\nonumber
\end{equation}
because the basic IL parameters, as we agreed, are unchanged. Handling the
'mixed' component average we have it in the form as reads
\begin{equation}
\label{15}
\langle G_{\mu\nu}^a(A,B)~G_{\mu\nu}^a(A,B)
\rangle_{\omega z}= \frat{18~\pi^2~\rho^2}{V}\frat{N_c}{N_c^{2}-1}
~B^{b}_\mu ~B^{b}_\mu~,
\end{equation}

Finally, collecting all appropriate terms we find the effective action for
the external field in IL as
\begin{equation}
\label{16} \langle \langle S \rangle\rangle_{\omega z}=\int dx~
\left( \frat{G(B)~G(B)}{4}+\frat{m^2}{2}~B^2\right)+N~\beta~,
\end{equation}
\begin{equation}
\label{17}
m^2=9\pi^2~n~\rho^2~\frat{N_c}{N_c^{2}-1}~,
\end{equation}
here $N$ is the full number of particles in volume $V$ with $n=N/V$ and
a single pseudo-particle action $\beta=8\pi^2/g^2$. The last term of
Eq.(\ref{16}) introduces the contribution of purely instanton component
$\langle G(A) G(A)\rangle_{\omega z}$. The contribution of repulsive term which
fixes the pseudo-particle size in IL is omitted in Eq.(\ref{16}) so long as it
is not a principal point in this context and adding it, leads to the
insignificant correction to the last condensate term in Eq.(\ref{16}). An amusing point is
that the mass term of Eq.(\ref{17}) has been well-known for rather long time and as
a matter of fact fixing the pseudo-particle size in the variational procedure of
Ref.\cite{DP} is provided just by this mechanism of mass generation. With the characteristic
IL parameters ($N_c=3$ and number of flavours $N_f=2$)
$n/\Lambda_{QCD}^4=1.2$, $\bar\rho \Lambda_{QCD}=0.27$, $\beta=18$ the mass estimate
$m \sim 0.44 {\mbox {GeV }}$ looks pretty encouraging for
$\bar\rho \sim 1 {\mbox {GeV }}$ and $\Lambda_{QCD}$ in the interval of
$200$ --- $300$  {\mbox {GeV}}. The compatibility conditions for the
equations resulting from Eq.(\ref{16}) is $\partial_\mu B_\mu = 0$ which is
satisfied by the pseudo-particle field Eq.(\ref{2}) as well.

Turning now to the next term of cluster decomposition to calculate the
effective Lagrangian corrections we conclude immediately that in the second cumulant
$$\frat12\left\langle\left\langle \int dx_x \frat{G~G}{4} \int d x_2
\frat{G_2~G_2}{4}\right\rangle\right\rangle~,$$
there are two nontrivial terms
\begin{eqnarray}
\label{18} &&\frat12\left\langle \int
dx_1~2~\frat{G_{\mu\nu}^a(B)~G_{\mu\nu}^a(A)}{4} \int
dx_2~2~\frat{G_{\alpha\beta}^b(B_2)~G_{\alpha\beta}^b(A_2)}{4}\right\rangle~,\\
\label{19} &&\frat12\left\langle \int
dx_1~2~\frat{G_{\mu\nu}^a(A)~G_{\mu\nu}^a(A,B)}{4} \int
dx_2~2~\frat{G_{\alpha\beta}^b(A_2)~G_{\alpha\beta}^b(A_2,B_2)}{4}\right\rangle~,
\end{eqnarray}
here the index $2$ underlines the fact that corresponding functions are
dependent on $x_2$. The remaining terms originate from either the interference terms (and
are cancelled by the contribution of the first cumulant squared) or lead to the
contributions anharmonic in $B$ which are not in our interest for this paper.
It was analyzed for the first time in Ref.\cite{CDG} that $G(B)G(A)$ in
(\ref{7}) generates the dipole interaction. However, this interaction does not manifest
itself in the first term of cluster decomposition if the averaging over the colour
orientation is performed. It comes into focus starting on the second order of
decomposing. In particular Eq.(\ref{18}) can be presented in the following form
\begin{eqnarray}
\label{20}
&&\frat12\left\langle \int dx_1~2~\frat{G_{\mu\nu}^a(B)~G_{\mu\nu}^a(A)}{4}
\int
dx_2~2~\frat{G_{\alpha\beta}^b(B_2)~G_{\alpha\beta}^b(A_2)}{4}\right\rangle_\omega=\nonumber\\
&&=\frat12~\frat{1}{N_c^{2}-1} \int d x_1~d x_2~\frat{G_{\mu\nu}^a(B)~
 G_{\alpha\beta}^b(B_2)}{4}~G_{\mu\nu}^b(A)~G_{\alpha\beta}^{b}(A_2)~,
\end{eqnarray}
if one ~exploits Eq.(\ref{5}) and Eq.(\ref{12}) keeping in mind
that $G_{\mu\nu}^b(A)~G_{\alpha\beta}^{b}(A_2)$ is colour
independent because of the identity
$\omega^{ab}\omega^{ac}=\delta^{bc}$. Eq.(\ref{20}) should be also
averaged over the pseudo-particle positions which results in the
correlation function for the instantons in singular gauge
developing the following form obtained in Ref.\cite{1} $$\int
\frat{dz}{V}~G_{\mu\nu}^a(A)~G_{\alpha\beta}^a(A_2)=
\frat{1}{V}~\frat{16}{g^2}~\left(\delta_{\mu\alpha}\delta_{\nu\beta}-
\delta_{\mu\beta}\delta_{\nu\alpha}+\varepsilon_{\mu\nu\alpha\beta}\right)~
I_s\left(\frat{|\Delta|}{\rho}\right)~, $$ where
$\Delta=|x_1-x_2|$, and for the anti-instanton the substitution
$\varepsilon\to-\varepsilon$ should be done. The analytical form
of function $I_s$ is not our priority here, however, it is shown
in Fig.1. If the numbers of instantons and anti-instantons are
balanced then the term proportional to the tensor $\varepsilon$
disappears and the contribution of correlator in IL is given by
$$\frat{N}{V}\int
dz~G_{\mu\nu}^a(A)~G_{\alpha\beta}^a(A_2)=\frat{16}{g^2}~n~
\left(\delta_{\mu\alpha}\delta_{\nu\beta}-
\delta_{\mu\beta}\delta_{\nu\alpha}\right)~
I_s\left(\frat{\Delta}{\rho}\right)~.$$

\begin{figure*}[!tbh]
\begin{center}
\includegraphics[width=0.75\textwidth]{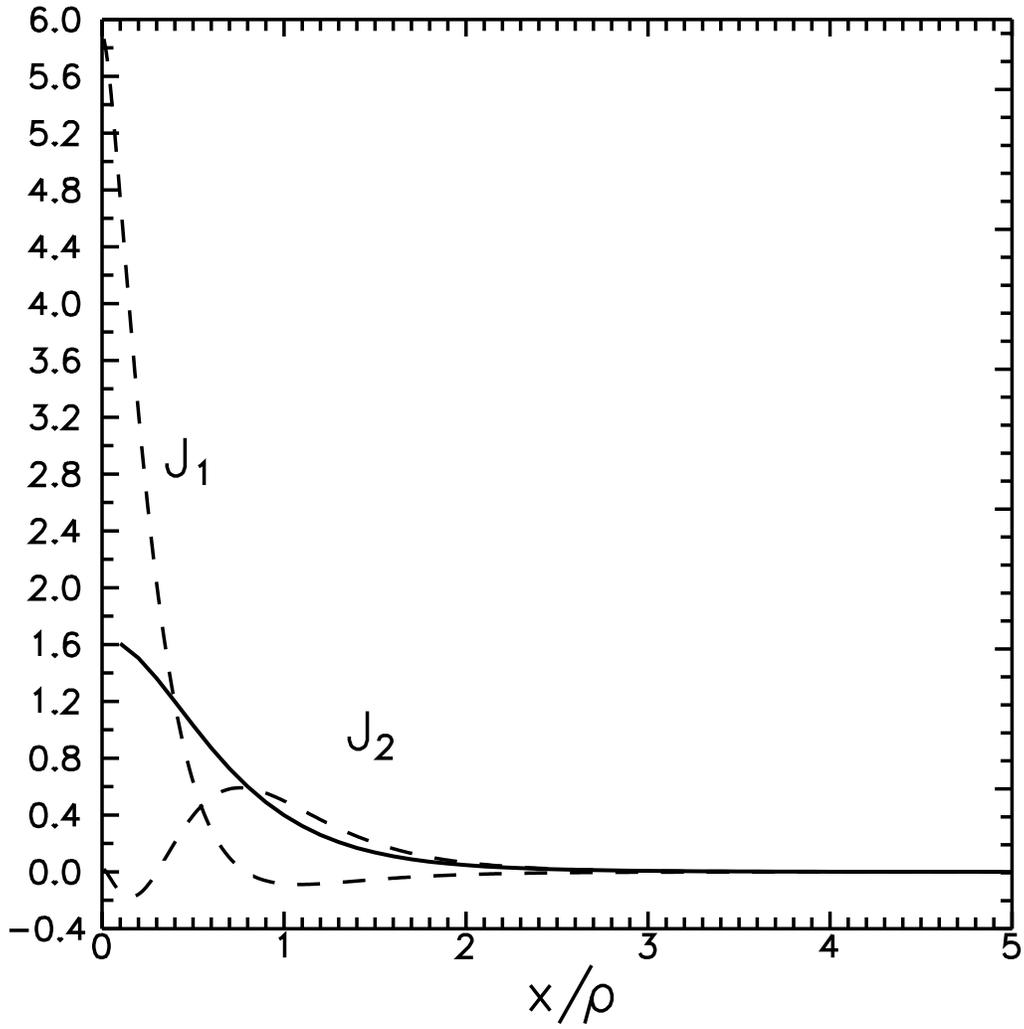}
\end{center}
  \vspace{-7mm}
 \caption{Correlation function $I_s$ is given by solid line and the correlation
functions $J_1$ and $J_2$ are given by the dashed lines.}
  \label{comp}
\end{figure*}

Now collecting the terms together we find the contribution of Eq.(\ref{18}) in
the IL approach as
$$\frat{16}{g^2}~\frat{1}{N_c^{2}-1}~n~
\int d x_1 d
x_2~I_s\left(\frat{\Delta}{\rho}\right)~G_{\mu\nu}^a(B)~
G_{\alpha\beta}^a(B_2)~. $$
Clearly, it leads to an abatement of initial action
and it is more convenient for analyzing to present the non-local factor of
dielectrical susceptibility type in the Fourier components Ref.\cite{CDG}
$$\int dk \left(1-\frat{16}{g^2}~\frat{1}{N_c^{2}-1}~ n~\widetilde
I_s(k\rho)\right) ~ G^a_{\mu\nu}[B(k)]~
G_{\alpha\beta}^a[B(-k)]~.$$
Numerical estimate of $\widetilde I_s(k\rho)$ at the zero value of argument is
$$\widetilde I_s(0)\sim 6~\rho^4~,$$
and at $N_c=3$, $N_f=2$ the correction coefficient can be estimated as
$\kappa=\frat{16}{g^2}~\frat{1}{N_c^{2}-1}~ n~\widetilde
I_s(0)\sim 0.013$.

Analyzing now the term Eq.(\ref{19}) we present it as
\begin{eqnarray}
\label{21} &&\frat12\left\langle \int
dx_1~2~\frat{G_{\mu\nu}^a(A)~G_{\mu\nu}^a(A,B)}{4} \int
dx_2~2~\frat{G_{\alpha\beta}^b(A_2)~G_{\alpha\beta}^b(A_2,B_2)}{4}\right\rangle_\omega=\nonumber\\
&&=\frat12~\left\langle\int d x_1~d
x_2~\omega^{ak}~G_{\mu\nu}^k(A)~f^{amn}~\omega^{nl}
(B^{m}_\mu~\bar\eta_{l\nu\gamma}-B^{m}_\nu~\bar\eta_{l\mu\gamma})~a_\gamma
\times\right.\nonumber\\ &&
\left.\times\omega^{bc}~G_{\alpha\beta}^k(A_2)~f^{bde}~\omega^{ef}
(B^{d}_{2\alpha}~\bar\eta_{f\beta\delta}-B^{d}_{2\beta}~\bar\eta_{f\alpha\delta})~a_{2\delta}
\right\rangle_\omega~.
\end{eqnarray}
and imply the dependence of $G$ on the colour matrix $\omega$ might be given by
the common factor (without introducing new symbol for $G$). Formally, this term
looks like the next one expanding in $1/N_c$, i.e. ($\sim \omega^4$). However, using the
identity for colour matrices
$$f^{man}~\omega^{ak}\omega^{nl}=\varepsilon^{klg}~\omega^{mg}~,~~$$
we have
\begin{equation}
\langle f^{man}~\omega^{ak}\omega^{nl}~f^{dbe}~\omega^{bc}\omega^{ef}\rangle=
\frat{\delta^{md}}{N_c^{2}-1}~
\left(\delta^{kc}\delta^{lf}-\delta^{kf}\delta^{lc}\right)~,\nonumber
\end{equation}
and then Eq.(\ref{21}) receives the following form
\begin{eqnarray}
&&\frat12\left\langle \int dx_1~2~\frat{G_{\mu\nu}^a(A)~G_{\mu\nu}^a(A,B)}{4}
\int
dx_2~2~\frat{G_{\alpha\beta}^b(A_2)~G_{\alpha\beta}^b(A_2,B_2)}{4}\right\rangle_\omega=\nonumber\\
 && =2~\frat{1}{N_c^{2}-1}~\int d x_1 d
x_2~\left[~G_{\mu\nu}^k(A)~G_{\alpha\beta}^k(A_2)~
\bar\eta_{l\nu\gamma}\bar\eta_{l\beta\delta}-G_{\mu\nu}^k(A)~G_{\alpha\beta}^l(A_2)~
\bar\eta_{k\beta\delta}\bar\eta_{l\nu\gamma}\right] ~a_\gamma
a_{2\delta} ~B^{m}_\mu B^{m}_{2\alpha}~,\nonumber
\end{eqnarray}
The lower line here develops this form because of an asymmetric property of
tensor
$G$. Averaging over the pseudo-particle positions we may extract the
correlation
function in the following form
\begin{eqnarray}
&& \int \frat{d z}{V}~\left[~G_{\mu\nu}^k(A)~G_{\alpha\beta}^k(A_2)~
\bar\eta_{l\nu\gamma}\bar\eta_{l\beta\delta}-
G_{\mu\nu}^k(A)~G_{\alpha\beta}^l(A_2)~
\bar\eta_{k\beta\delta}\bar\eta_{l\nu\gamma}\right] ~a_\gamma
a_{2\delta}=\nonumber\\
&&=\frat{16}{g^2}~\frat{1}{V}~\left[J_1\left(\frat{\Delta}{\rho}\right)~\delta_{\mu\alpha}+
J_2\left(\frat{\Delta}{\rho}\right)~\hat\Delta_\mu
\hat\Delta_\alpha\right]~,\nonumber
\end{eqnarray}
where $\hat\Delta=\frac{x_2-x_1}{|x_2-x_1|}$ is the unity vector.

The simple algebra allows us to calculate the functions
\begin{eqnarray}
&&J_1=\int dy \frat{\rho^4}{(y^2+\rho^2)^3}~
\frat{\rho^4}{(z^2+\rho^2)^3}~\frat{1}{|y|}
~\frat{1}{|z|}~\frat13~(16~t^3-8~t+4~pq+6~(p^2+q^2)~t-12~t^2~pq)~,\nonumber\\
 &&J_2=\int dy
\frat{\rho^4}{(y^2+\rho^2)^3}~
\frat{\rho^4}{(z^2+\rho^2)^3}~\frat{1}{|y|}
~\frat{1}{|z|}~\frat43~(4~t^3+5~t-4~pq-6~(p^2+q^2)~t+12~t^2~pq)~,\nonumber
\end{eqnarray}
with $z=y+\Delta$, $t=\frac{(y~ z)}{|y||z|}$,
$p=\frac{(y~\Delta)}{|y||\Delta|}$,
$q=\frac{(z~\Delta)}{|z||\Delta|}$. Similarly to $I_s$ we do not need their
explicit forms here but one may estimate their behaviours looking at the dashed lines in
Fig.1. Finally, the additional contribution to the mass term reads as
$$\frat{1}{N_c^{2}-1}~\frat{32}{g^2}~n~\int dx_1d x_2
\left[J_1\left(\frat{\Delta}{\rho}\right)~\delta_{\mu\alpha}+
J_2\left(\frat{\Delta}{\rho}\right)~\hat\Delta_\mu
\hat\Delta_\alpha \right] ~B^a_{\mu} B^a_{2\alpha}~,$$
and in the Fourier components as
\begin{equation}
\label{22}
\int d k~\left[ \frat{m^2}{2}-\frat{32}{g^2}~\frat{1}{N_c^{2}-1}~n~\left(
\widetilde J_1(k\rho)~\delta_{\mu\alpha}+\widetilde J_2(k\rho)~\hat k_\mu \hat
k_\alpha \right) \right]~B^a_{\mu}(k) B^{a}_{\alpha}(-k)~.
\end{equation}
Estimating numerically the nonlocal correction to the mass we find out that
$\widetilde J_1(0)\sim -1.4 ~\rho^2$ unlikely above result. Then the mass term
and corresponding correction in Eq.(\ref{22}) come about at zero momentum
$9\pi^2~N_c$ and $(-\frac{4\beta}{\pi^2}~1.4)$, respectively. At the
characteristic value $\beta\sim 18$ it means the quantitative correction smallness or,
globally, the corrections initiated by the second term of cumulant expansion are negligible
at the contemporary values of basic IL parameters.
There is another contribution to the effective Lagrangian which comes
from the interaction of sources generating the external field with
(anti-)instanton superposition
$$S_{int}=\sum_{i=1}^N\int~dx~j^a_{\mu}(x)A^a_{\mu}(x;\gamma_i)~.$$
Making use the cluster decomposition one expects the possibility to calculate
corresponding small contributions (if the sources are treated in the quasiclassical
approximation) which are given by the correlation functions of the form
$\langle A^a_{\mu}(x;\gamma) A^b_{\nu}(y;\gamma)\rangle_\gamma$ Ref.\cite{CDG}.

As the conclusion of this effective Lagrangian analysis Eq.(\ref{16}) it is practical
to address another approach to the problem Ref.\cite{cnf}. Let us suppose the quasi-classical
field $B$ is described in the infra-red momentum region by the initial Yang-Mills
action without the term breaking down gauge symmetry as before. In particular, we
consider the field of point-like Euclidean source of intensity $e$ with only one
non-zero $n$-th component
$$B^a_{\mu}(x)=({\vf
0},\delta^{an}~\varphi),~\varphi=\frat{e}{4\pi}~\frat{1}{|{\vf
x}|}~. $$
Then $B^2$ integrated over the 4-dimensional space gives
$$\int dx \left(\frat{e}{4\pi~|{\vf
x}|}\right)^2=\frat{e^2}{4\pi}~X_4~L~,$$
where $X_4$, $L$ are some formal upper
limits of corresponding integrals. In this approach the contribution of the first
cumulant Eq.(\ref{16}) could be written down as
\begin{equation}
\label{23}
\langle\langle S\rangle\rangle_{\omega z}=E~X_4~,~~
E=\frat{e^2}{4~\pi}~\frat{1}{r_0}+\sigma~L+\beta~n~L^3~,
\end{equation}
with $\sigma=\frac{9\pi}{8}\frac{N_c}{N_c^{2}-1}~e^2 ~n{ \overline {\rho^2}}$.
The first term in defining $E$ comes from the Coulomb energy of point-like
source and $r_0$ represents a formal particle radius. The last term is originated by
the gluon condensate and the previous term looks like negligibly small correction to
the condensate term. However, this contribution linearly increasing with $L$ is
proportional to $e^2$ and has different physical meaning as a term additional
to the self-energy of source. In other words, it demonstrates an impossibility for the
source with an open colour to be available in IL because the amplitude of
such a state is very strongly suppressed ($e^{-S}$) comparing to the condensate contribution if
the screening effects are not taken into account. For the dipole in 'isosinglet'
($s$) and 'isotriplet' ($t$) states (i.e. $N_c=2$) we obtain
$$B^a_{\mu}(x)=({\vf 0},\delta^{a3}~\varphi),~\varphi=\frat{e}{4\pi}~\left(\frat{1}
{|{\vf x}-{\vf z}_1|} \mp \frat{1}{|{\vf x}-{\vf z}_2|}\right)~,\nonumber $$
where ${\vf z}_1$, ${\vf z}_2$ are the dipole coordinates what leads to
$$\int dx~  B_{s}^2=\frat{e^2}{4\pi}~X_4~l~,~~~\int dx~
B_{t}^2=\frat{e^2}{4\pi}~X_4~(4~L-l)~,$$
with $l= |{\vf z}_1-{\vf z}_2|$ to be the distance separating ~sources. We have
another confirmation of suppression effect for the states with open colour in IL,
i.e. the energy of 'isosinglet' dipole state increases with $l$ enlarging and the
corresponding coefficient is $\sigma \sim 0.6 {\mbox{ GeV/fm}}$ if we take
$e\sim g$.

Thus, we are quite allowed  to conclude the regime of weak external field in IL is
described by effective Lagrangian Eq.(\ref{16}) and basic IL
parameters are  within a well adapted interval. Moreover all the corrections originated by the second
cumulant should be certainly neglected.

Obviously, this conclusion will be considerably strengthened if a criterion
of external field weakness is well defined and the point of how crucial is an
assumption of the IL parameters unchanged is clarified. We are going to modify slightly the
variational procedure of Ref.\cite{DP} to implement  possibility of the
changing IL parameters. We retain
here the same designations to demonstrate precisely where the changes are introduced and
imply $S(B,\gamma)$ in Eq.(\ref{9}) in the following form
\begin{equation}
\label{24}
S(B,\gamma)=-\sum \ln d(\rho_i)+ \beta~U_{int}+\sum U_{ext}(\gamma_i,B)+S(B)~,
\end{equation}

The first term here describes one-instanton contributions with the following
distribution function over the (anti-)instanton sizes
\begin{equation}
\label{25 } d(\rho)= C_{N_c} \Lambda_{QCD}^b~\rho^{b-5}
\widetilde\beta^{2 N_c},
\end{equation}
where $b=\frac{11}{3} N_c-\frac{2}{3}N_f$, $\widetilde\beta=-b \ln(\Lambda_{QCD}~
\bar\rho)$,
$$C_{N_c}\approx\frac{4.66~\exp(-1.68 N_c)}{\pi^2(N_c-1)!(N_c-2)!}~.$$
The second term of Eq.(\ref{24}) is responsible for providing pseudo-particles
with repulsive interaction which fixes their sizes. The characteristic single
instanton action is defined on the scale of average pseudo-particle size
$\beta=\beta(\bar\rho)$
where $\beta(\rho)=-\ln C_{N_c}-b \ln(\Lambda_{QCD}~ \rho)$. The partial
pseudo-particle
contributions grouped in the third term and we take only
$$U_{ext}(\gamma_i,B)=\int dx ~\frat{G_{\mu\nu}^a(A_i,B)~
G_{\mu\nu}^a(A_i,B)}{4}~, $$
because the other contributions at the standard IL parameters are small as we
have seen. At last, the fourth term represents simply the Yang-Mills action of the $B$
field
$$S(B)=\int dx~\frat{G_{\mu\nu}^a(B)~ G_{\mu\nu}^a(B)}{4}~. $$

The well-known property of exponential makes it possible to estimate the
generating functional of Eq.(\ref{9}) with the approximating functional as
\begin{equation}
\label{26} Y\ge Y_1~\exp(-\langle S-S_1\rangle)~,
\end{equation}
where
$$Y_1=\int D[B]~ \frat{1}{N!} \int \prod_{i=1}^N~
d\gamma_i~~e^{-S_1(B,\gamma)-S(B)}~,
~~~S_1(B,\gamma)=-\sum \ln \mu(\rho_i)~, $$
and $ \mu(\rho)$ is an effective one-particle distribution function which may
be derived with the variational procedure. In our particular situation a mean value of
corresponding difference is given by
\begin{eqnarray}
\label{27}
&&\langle S-S_1\rangle=\frat{1}{Y_1}~\frat{1}{N!}
\int~\prod_{i=1}^N~ d
\gamma_i~ [\beta~U_{int}+U_{ext}(\gamma,B)-\sum\ln d(\rho_i)+\sum \ln
\mu(\rho_i)]
~e^{~\sum \ln \mu (\rho_i)}=\nonumber\\
&&=\frat{N}{\mu_0}~\int d \rho~ \mu
(\rho)~\ln \frat{\mu (\rho)}{d
(\rho)}+\frat{\beta}{2}~\frat{N^2}{V^2}~\frat{1}{\mu_0^{2}} ~\int d\gamma_1
d\gamma_2~U_{int}(\gamma_1,\gamma_2)~ \mu (\rho_1)
 \mu (\rho_2)+\nonumber\\
&&+\int dx~ \frat{N}{V}~\frat{1}{\mu_0}~\int d \rho~\mu (\rho)~\rho^2
 \zeta~B^2=\nonumber\\
&&=\int dx~ n~ \left( \frat{1}{\mu_0}~\int d \rho~ \mu (\rho)~\ln
\frat{\mu (\rho)}{d (\rho)}+\frat{\beta \xi^2}{2}~n \left(
\overline{ \rho^2}\right)^2 +\zeta \overline{ \rho^2}~B^2\right)~,
\end{eqnarray}
with $\zeta=\frac{9~\pi^2}{2}~\frac{N_c}{N_c^2-1}$,
$\xi^2=\frac{27}{4}\frac{N_c}{N_c^{2}-1} \pi^2$, $\mu_0=\int d
\rho~ \mu (\rho)$. Here we estimate the functional in the adiabatic (long
wave-length) approximation. It means we consider the IL elements of some characteristic
size (of the same order of magnitude as the mean distance between pseudo-particles) being
equilibrated by the presence of some fixed field $B$. Then calculating the optimal
configurations of pseudo-particles we found out the effective action in the mean field.
Eq.(\ref{27}) is given just in the form underlining that an integration is performed over liquid
elements and the proper parameters describing their states could be dependent on the
external field, i.e. could be the functions of coordinate $x$. Physical meaning of such a
functional is quite transparent, it implies that each separate element of IL possesses a
characteristic aptitude of screening external field assessed by $U_{ext}$.

\begin{figure*}[!tbh]
\begin{center}
\includegraphics[width=0.75\textwidth]{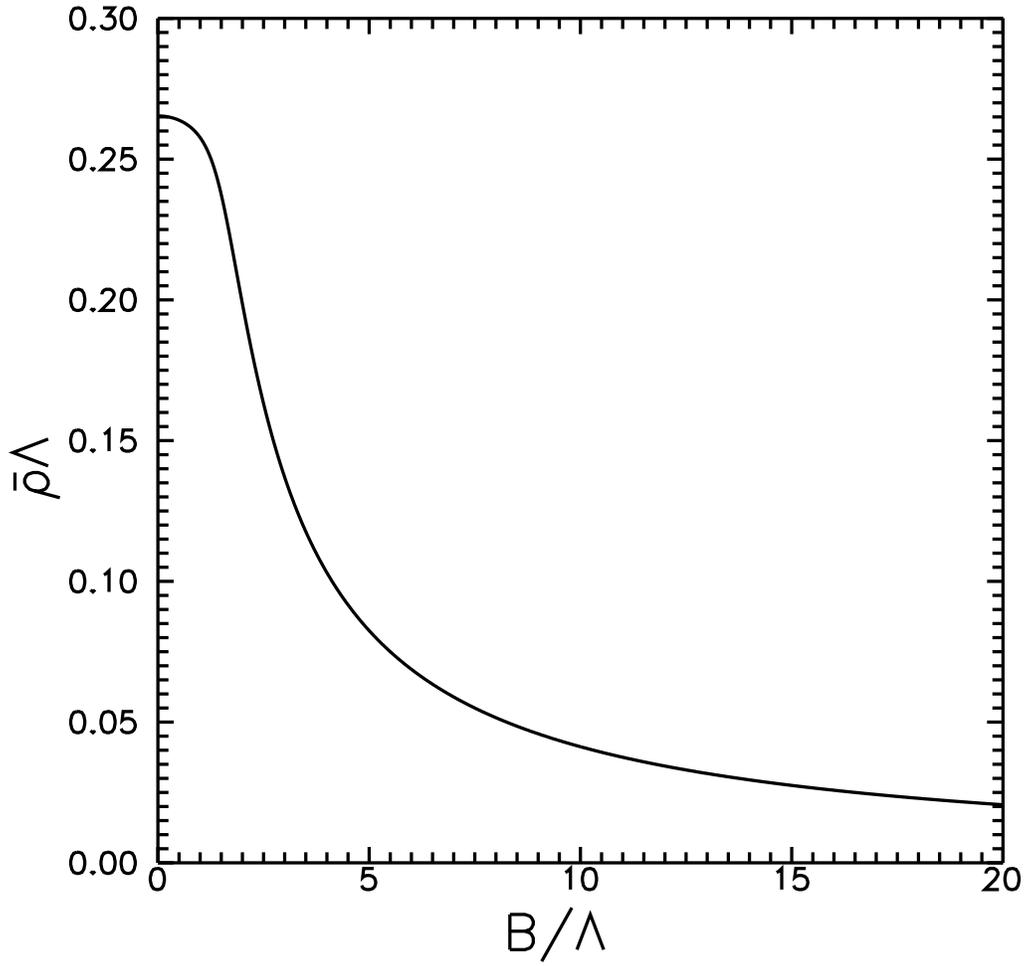}
\end{center}
  \vspace{-7mm}
 \caption{Pseudo-particle mean size as a function of applied external field.}
  \label{rho}
\end{figure*}

Calculating the variation of $\langle S-S_1\rangle$ in $\mu (\rho)$ we have
$$\mu (\rho)= C~ d (\rho) ~e^{-(n\xi^2\overline{\rho^2}+\zeta~ B^2)\rho^2}~,
$$
where $C$ is an arbitrary constant and we fix it demanding the coincidence of
its value when the external field is absent with its vacuum average. Then
\begin{equation}
\label{28}
\mu (\rho)= C_{N_c} \widetilde \beta^{2N_c}\Lambda_{QCD}^b \rho^{b-5}
 ~e^{-(n\xi^2\overline{\rho^2}+\zeta~ B^2)\rho^2}~.
\end{equation}
 and making use the definition of an average as
$$\overline{\rho^2}=\frat{\int d \rho~ \rho^2~\mu (\rho)}{\mu_0}~,
$$
we obtain the practical relation between mean pseudo-particle size and the IL
density
\begin{equation}
\label{29}
(n~\beta~\xi^2~\overline{\rho^2}+\zeta~B^2)~\overline{\rho^2}\simeq
\nu~,
\end{equation}
where $\nu=\frat{b-4}{2}$. Apparently, it results in a well-known form of
pseudo-particle
size distribution
\begin{equation}
\label{30}
\mu (\rho)= C_{N_c} \widetilde \beta^{2N_c}\Lambda_{QCD}^b \rho^{b-5}
 ~e^{-\nu~\frac{\rho^2}{\overline{\rho^2}}}~.
\end{equation}
Now Eq.(\ref{29}) allows us to formulate the criterion we are interested in. It
looks like $\zeta B^2 \ll n~\beta~\xi^2~\overline{\rho^2}$ and for the IL parameters
mentioned above it is $B \ll 400$ MeV.

\begin{figure*}[!tbh]
\begin{center}
\includegraphics[width=0.75\textwidth]{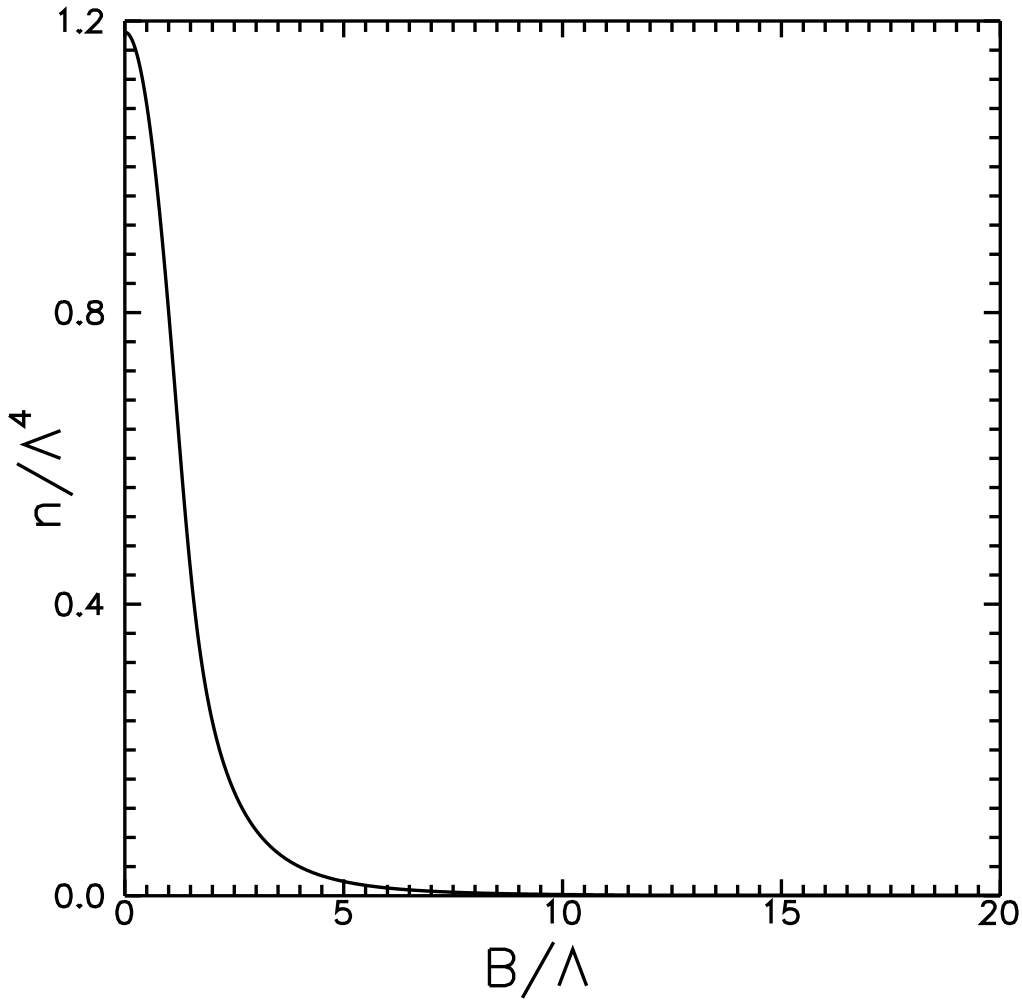}
\end{center}
  \vspace{-7mm}
 \caption{The IL density as a function of applied external field.}
  \label{n}
\end{figure*}

Dealing with Eq.(\ref{27}) and Eq.(\ref{30}) the generating functional estimate
Eq.(\ref{26}) may be presented as
\begin{equation}
\label{31}
Y\ge \int D[B]~e^{-S(B)}~e^{-F}~,
\end{equation}
$$F=\int dx~
n~\left\{\ln\frat{n}{\Lambda_{QCD}^4}-1-\frat{\nu}{2}+\frat{\zeta~\overline{\rho^2}~B^2}{2}-
\ln [\Gamma(\nu)~C_{N_c}~\widetilde \beta^{2N_c}]-\nu~\ln
\frat{\overline{\rho^2}}{\nu} \right\}~. $$ Making use of the relation
Eq.(\ref{29})
it is not difficult to find the maximum of functional Eq.(\ref{31}) in the
IL parameters at the fixed $B$ value as a solution of transcendental equation
($\frac{d F}{d\bar\rho}=0$). As an information we give the simple
expression of its derivative in $n$
$$F'_{n}=\ln\frat{n}{\Lambda_{QCD}^4}+\frat14~\frat{n^2\xi^4\beta~b~(\overline{\rho^2})^3}
{2n\beta~\xi^2~ \overline{\rho^2}~ \zeta~B^2-
n~\xi^2~\frac{b}{2}~\overline{\rho^2}}-
 \ln [\Gamma(\nu)~C_{N_c}~\widetilde \beta^{2N_c}]-2N_c~n~\frat{\widetilde
\beta'_{n}}
{\widetilde \beta}-\nu~\ln \frat{\overline{\rho^2}}{\nu}~.$$
Fig.2 and Fig.3 demonstrate the solutions for $\bar\rho$ and $n$ at $N_c=3$ and
$N_f=2$ as the functions of field $B$. Fig.4 shows the plot of free energy density
$f/\Lambda_{QCD}^4$ where $F=\int dx~ f$ and convinces IL is steady as to an impact
of external field. At strong external field the IL parameters are given by
the following asymptotic formulae
\begin{eqnarray}
&&
\overline{\rho^2}\simeq\frat{\nu}{\zeta~B^2}~\left(1-\frat{n~\nu~\beta\xi^2}{\zeta^2~B^4}\right)~,\nonumber\\
&&
n \simeq \frat{\Gamma(\nu)~C_{N_c}~\widetilde \beta^{2N_c}
}{(\zeta~B^2)^{\nu}}~
\left(1+ \frat{\Gamma(\nu)~C_{N_c}~\widetilde \beta^{2N_c}
}{(\zeta~B^2)^{\nu}}~
\frat{N_c~b~\nu~\beta~\xi^2}{\zeta^2~B^4}\right)~.\nonumber
\end{eqnarray}
This regime starts somewhere around $B\Lambda_{QCD}^{-1}\sim 10$ at all the plots
given.

\begin{figure*}[!tbh]
\begin{center}
\includegraphics[width=0.75\textwidth]{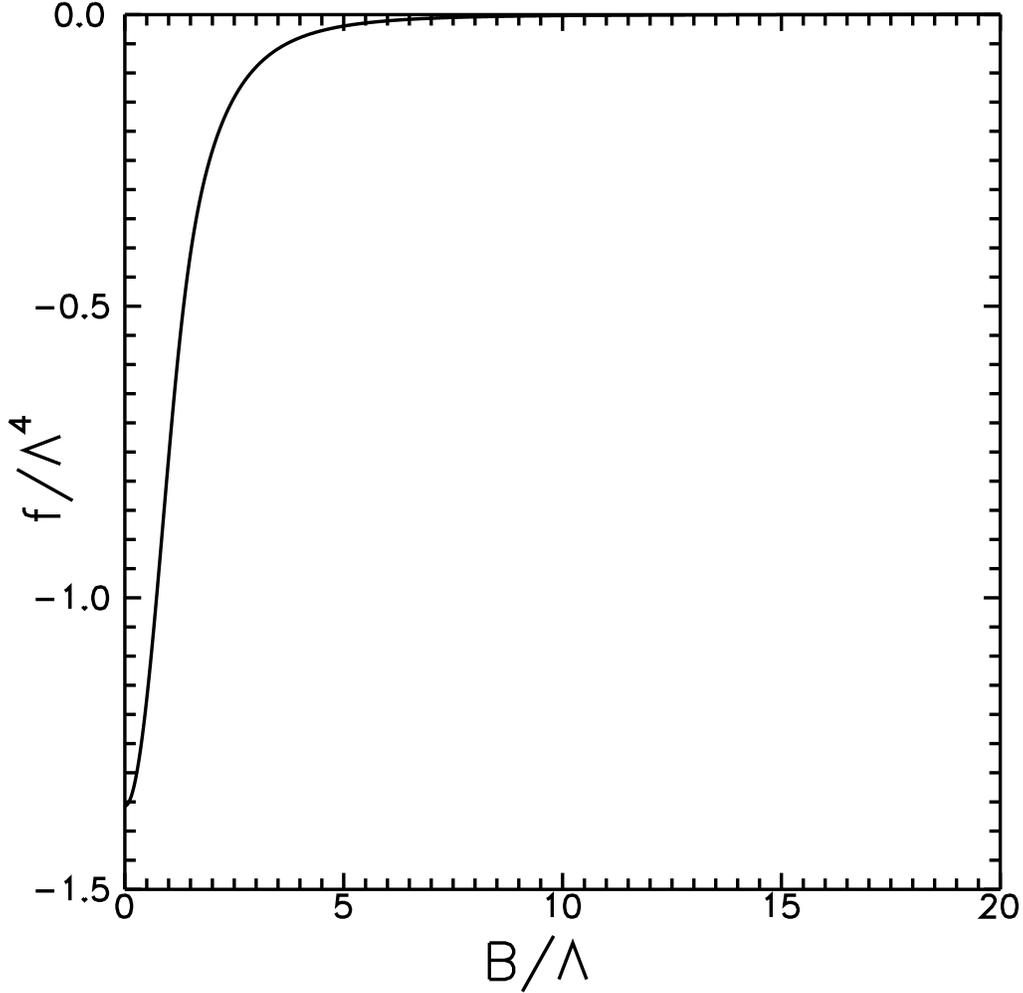}
\end{center}
  \vspace{-7mm}
 \caption{Free energy density as a function of external field $B$.}
  \label{free}
\end{figure*}

Thus, the effective action for the $B$ field is given by the following
nonlinear functional
\begin{equation}
\label{32}
S_{eff}=\int dx \left( \frat{G_{\mu\nu}^a(B)~
G_{\mu\nu}^a(B)}{4}+f[B] \right)~.
\end{equation}
This functional makes possible to calculate the external field as a function of $x$
and IL parameters $\bar\rho[B]$ and $n[B]$.

 To get any estimate of the IL feedback on the presence of external
field could be very practical for instanton liquid model. If so
let us try to extract such an estimate from very simple example. Now we will
search the minimum of effective action resolving the following boundary value problem
\begin{eqnarray}
\label{33} &&\triangle_r B=\frat{d f[B]}{d B}~,\\
&&B|_{r=r_0}=p(e)~,~~~\nabla_r B
\left|_{r=r_0}\right.=-\frat{e}{4\pi}\frat{1}{r_0^{2}}~.\nonumber
\end{eqnarray}
The source intensity here is controlled by $e$, and parameter $r_0$ sets
a radius of colour ball which we take as $\sim 0.1 \bar\rho$ (albeit it
is unessential) in order to avoid the difficulties in resolving the singular
boundary value problem of Eq.(\ref{33}). The solution could be accomplished
numerically probing such values of potential $p(e)$ which provide with
the solution going to zero magnitude at large values of $r$.

\begin{figure*}[!tbh]
\begin{center}
\includegraphics[width=0.75\textwidth]{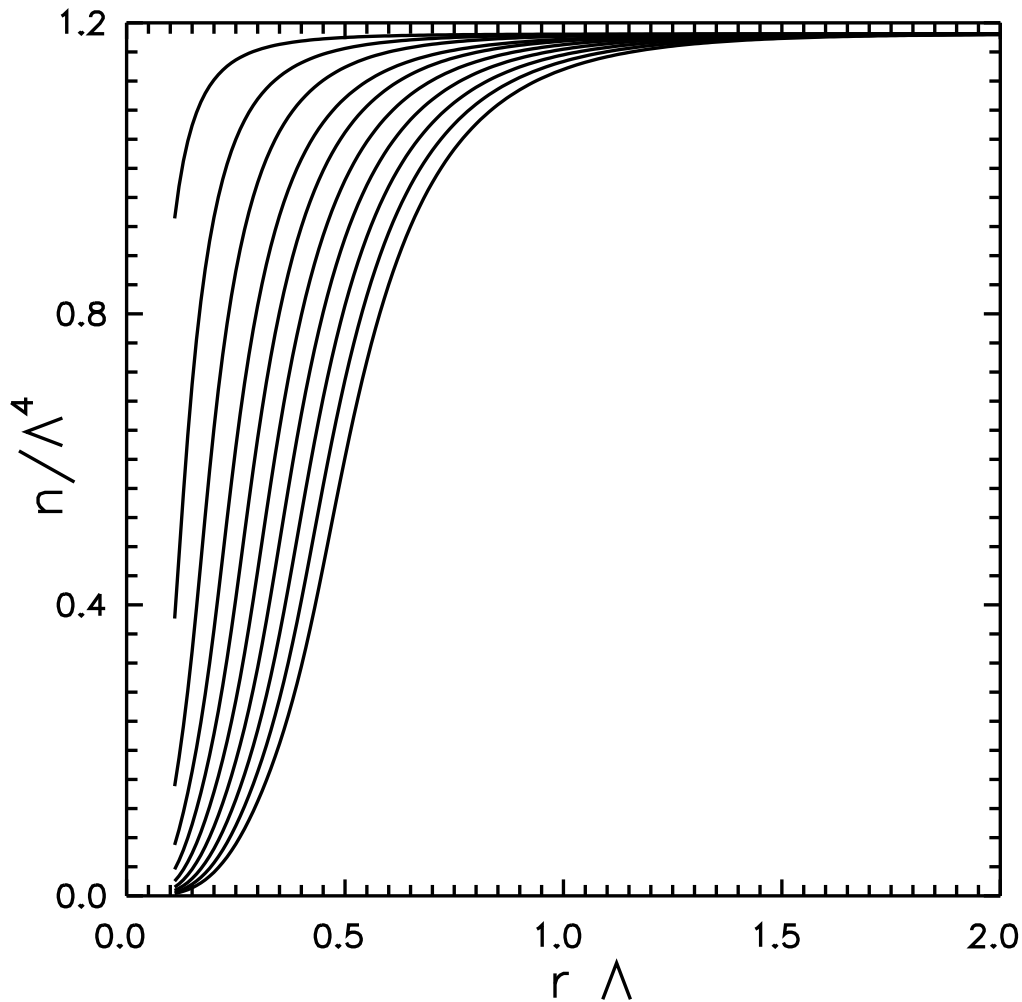}
\end{center}
  \vspace{-7mm}
 \caption{The IL density a a function of $r$ for different values of
intensity. Extreme right hand side line corresponds to $e/4\pi=1$. Going
to the left corresponds to changing $e/4\pi$ with a pace of $0.1$ up to
$e/4\pi=0.1$ what corresponds to extreme left hand side line.}
  \label{dens}
\end{figure*}

The IL density as a function of $r$ is plotted in Fig.5 for ten various
quantities of intensity. The extreme right hand side line corresponds to
$e/4\pi=1$ and the extreme left hand side corresponds to $e/4\pi=0.1$. The
same quantity of spacing corresponds to the lines running to the right
with intensity increasing. As it was expected the solution has the Yukawa like behaviour
which is well seen in Fig.6 where  $\ln(Br)$ is plotted as a function of
$r$ for four various values of intensity with the pace of $0.1$ and $e/4\pi=1$ for
the upper line.  Fitting it with the linear function gives the estimate of
screening radius which looks as follows
$$R_d\sim (1.24~ \Lambda_{QCD})^{-1}~,$$
Amazingly, this  results remains practically unchanged for the whole interval of the
intensities from $e/4\pi=0.1$ to $e/4\pi=1$ and implies that such a parameter
characterizes (at least in this interval of values) the screening properties of
IL itself. In a context of the model it looks like
rather soft scale for the screening radius and might be taken as another confirmation
of adiabatic approximation relevance for the Coulomb external field.
Besides it hints the instanton vacuum could provide the significant energy loss
of a parton at the later time of collision process contributing (and may be essentially)
to jet quenching.

Another point which is not studied here but worth of mentioning concerns the
manifestation of instanton ensemble screening properties. It turns out rather
unexpected because saturating instanton configuration is randomly oriented in colour space
both with external field and with it switched off. Actually, anisotropy in colour space
is routed (and is playing a role analogous to distribution function of colour charge)
in the gluon field action of corresponding exponent, namely in the 'mixed component of gluon field.
It is intuitively clear other instanton-like solutions
(sensitive to the presence of external colour field)
could be even more adequate configurations. Such solutions were investigated in
Ref.\cite{we1} and was demonstrated that suitable scale and proper
configurations affected by external field (crumpled instantons) could appear, indeed.

\begin{figure*}[!tbh]
\begin{center}
\includegraphics[width=0.75\textwidth]{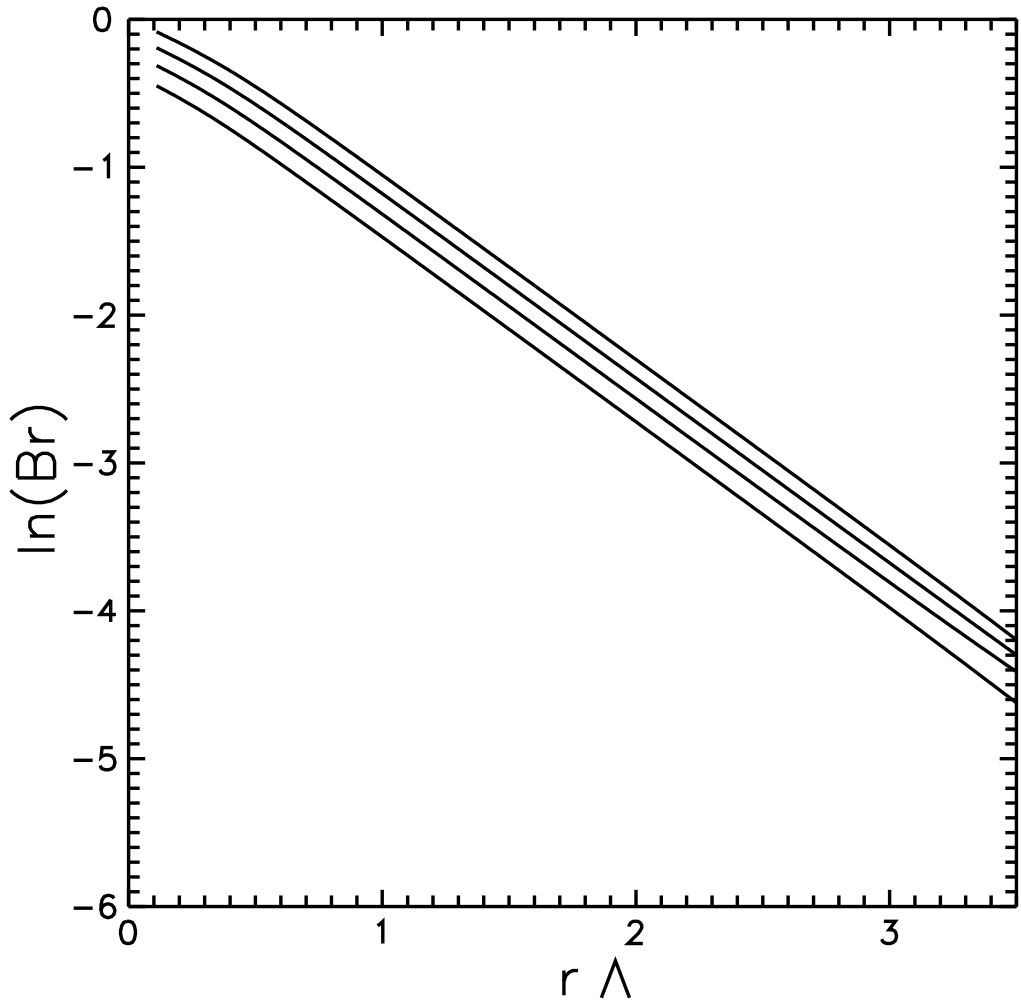}
\end{center}
  \vspace{-7mm}
 \caption{ $ln (Br)$ as a function of $r$ for four various solutions. The upper
line corresponds to $e/4\pi=1$. Going down the lines correspond to
decreasing $e/4\pi$ with spacing $0.1$.}
\label{exp}
\end{figure*}

Eventually let us comment  on how it is essential that we are dealing with singular
(anti-)instanton ensemble as a saturating configuration. Apparently, the screening
properties of effective Lagrangian for external field $B$ could be provided by
any stochastic configuration of small characteristic size. The assumption
of superposition ansatz  validity occurs crucial to have all the leading contributions coming
from the 'mixed' (repulsive) component  of $G(A,B)$ again. Another solution of the
problem may appear, of course, in the quantum approach but this discussion is out of this
paper scope. Studying the pseudo-particle behaviour while  inside
(anti-)instanton medium ($n\neq 0$) one could explore the interrelation of two mechanisms
(the repulsive interaction and freezing the coupling constant out Ref.\cite{Sh}) of fixing instanton size.

The paper was supported by INTAS-04-84-398,
NATO PDD(CP)-NUKR980668.
The authors are also indebted to  Barbara Jacak and Tetsuo Hatsuda for
stimulating discussions.

\end{document}